# Supra-Threshold Control of Peripheral LOD

Benjamin Watson*  
Northwestern University

Neff Walker[+]  
UNICEF

Larry F. Hodges[°]  
UNC-Charlotte

## Abstract

Level of detail (LOD) is widely used to control visual feedback in interactive applications. LOD control is typically based on perception at threshold – the conditions in which a stimulus first becomes perceivable. Yet most LOD manipulations are quite perceivable and occur well above threshold. Moreover, research shows that supra-threshold perception differs drastically from perception at threshold. In that case, should supra-threshold LOD control also differ from LOD control at threshold?

In two experiments, we examine supra-threshold LOD control in the visual periphery and find that indeed, it should differ drastically from LOD control at threshold. Specifically, we find that LOD must support a task-dependent level of reliable perceptibility. Above that level, perceptibility of LOD control manipulations should be minimized, and detail contrast is a better predictor of perceptibility than detail size. Below that level, perceptibility must be maximized, and LOD should be *improved* as eccentricity rises or contrast drops. This directly contradicts prevailing threshold-based LOD control schemes, and strongly suggests a reexamination of LOD control for foveal display.

**CR Categories:** I.3.3 [Computing Methodologies]: Computer Graphics – Display Algorithms; I.3.7 [Computing Methodologies]: Computer Graphics – Virtual Reality

## 1 Introduction

Level of detail (LOD) [Luebke et al. 2003] allows control of the tradeoff between visual detail and interactivity, and is now in widespread use. Advanced LOD schemes [Luebke and Hallen 2001; Reddy 1998] implement strict models of the threshold conditions in which objects become perceivable and use them to ensure that any detail removed is unperceivable (sub-threshold). However in interactive settings, this *sub-threshold LOD control* has not allowed significant improvements in rendering speed. More typically, LOD control uses heuristics based only loosely on perceptual thresholds. These *threshold-based control* schemes are then used even when the detail being manipulated is completely perceivable (supra-threshold), achieving useful rendering speed gains. Yet perceptual research finds drastic differences between perception at and above threshold. If this is so, should supra-threshold LOD control also differ from threshold-based control?

We address this question using two experiments implementing LOD control in the visual periphery. Thresholds rise significantly in the periphery of the visual field [Kelly 1984; Koenderink et al. 1978], and most threshold-based LOD control schemes exploit this by lowering LOD as eccentricity (angular distance from the fovea) rises. Supra-threshold perceptual research offers little support for this approach, and yet we have found success using just such a scheme [1997a, 1997b]. We begin examining this apparent contradiction with a survey of related work. We then describe our experiments and discuss the results.

## 2 LOD and Visual Sensitivity

Research on visual sensitivity has focused primarily on sensitivity thresholds. These respond strongly to many parameters, most importantly contrast and spatial frequency in a manner described by the contrast sensitivity function (CSF) [Graham 1989]. The CSF at the fovea has a band-pass form, with sensitivity to contrast being relatively limited at the lowest spatial frequencies, peaking at roughly 5 cycles per degree (c/deg), and then dropping steeply to zero at roughly 60 c/deg (meaning visual acuity is approximately one minute of visual angle). Visual sensitivity also drops steeply as stimuli move into the visual periphery, with the corresponding CSF quickly taking on a lower sensitivity, low-pass form, and acuity dropping from 60 c/deg at the fovea to 24 c/deg at 6 degrees of eccentricity and eventually 2.7 c/deg at 50 degrees.

When stimuli have more than one spatial component (e.g. two frequencies), their thresholds are lower than the thresholds of any of their components. The most widely accepted model of visual function holds that the early stage of the visual system is made of many distinct visual channels or analyzers [Graham 1989], each sensitive to a different range of spatial frequencies, orientations and positions. When a complex stimulus is made up of spatial components clustered closely enough to generate a response in one channel, that response is proportional to the sum of the stimulus components. Thus when those components are all sub-threshold, the threshold of the complex stimulus is effectively lower than the threshold of any one of its components. This same *summing* explains masking, which raises the thresholds for perception of any one component in a spatially clustered complex stimulus. Interestingly, when the components of a complex stimulus are spatially distant and generate a response in many channels, summing still occurs, but it is sub-linear, with an exponent of roughly $\beta = 4$ in the Minkowski sum $(\Sigma R_i^\beta)^{1/\beta}$, where $R_i$ is the response of channel $i$ to the stimulus. Linear summing would then occur with an exponent $\beta = 1$ [Graham 1989; Bonneh and Sagi 1998, 1999]. Sub-threshold LOD control can probably safely ignore sub-threshold summing when estimating perceptibility, since control generally manages detail component by component rather than object by object.

Many LOD and display researchers have created systems either inspired by or explicitly modeled on threshold visual sensitivity [Parkhurst and Niebur 2002]. Howlett [1992], Yoshida et al. [1995] and CAE Electronics [Barrette 1986; Fernie 1995] have all proposed threshold-based head-mounted displays with high resolution central display insets surrounded by lower resolution display regions. Funkhouser and Sequin [1993] and Ohshima et al. [1996] have both proposed threshold-based LOD control systems that reduce detail when spatial frequencies or eccentricities were high. More recently, Luebke and Hallen [2001] as well as Reddy [1998] have designed interactive systems incorporating sub-threshold LOD control, using explicit models of the CSF and the effects of eccentricity to ensure that only sub-threshold, unperceivable detail is eliminated. Both Luebke and Hallen and Reddy found that the resulting gains in rendering speed were limited. Several other researchers [Bolin and Meyer 1998; Ramasubramanian et al. 1999; Volevich et al. 2000] have designed global illumination systems that used a similar sub-threshold approach, employing image comparison metrics based on the CSF and other higher-level models of visual sensitivity at threshold to give perceptual guidance to adaptive rendering. These systems resulted in a more significant improvement in rendering speed, perhaps because the detail being controlled in these non-interactive systems was fine enough to make discarding sub-threshold detail a meaningful optimization.

*watson@northwestern.edu, [+]nwalker@unicef.org, [°]lfhodges@uncc.edu

Many of the above LOD and display systems are threshold-based, likely because a purely sub-threshold approach brings only limited improvements in interactive rendering speed. We [1997a, 1997b] evaluated the effectiveness of threshold-based supra-threshold LOD control in high contrast settings using visual search tasks that demanded extensive use of the visual periphery. These tasks are a very basic and demanding element of most interactive applications, particularly those using wide field of view (FOV) displays. Our system preserved a full LOD central display region, and eliminated any detail in the display periphery above a fixed spatial frequency. We found that reducing LOD in this manner did not significantly impact search performance, even when all frequencies above 0.27 c/deg were eliminated from display regions beyond 15 deg eccentricity.

Confusingly, perceptual research on supra-threshold visual sensitivity calls the basic assumptions of threshold-based control into question. While more sparse than research on thresholds, supra-threshold research has consistently observed the phenomenon of *contrast constancy*, in which equal physical contrast results in equal perceived contrast, regardless of variation in spatial frequency or eccentricity [Blakemore et al. 1973; Cannon 1985; Georgeson and Sullivan 1975; Kulikowski 1976; Peli et al. 1996]. In the fovea, sensitivity to contrast is constant over all spatial frequencies when it is over roughly 10%. In the visual periphery, contrast sensitivity quickly becomes constant as it rises above threshold at each eccentricity. For example at 30 degrees of eccentricity, contrast sensitivity is constant over all spatial frequencies below 4 c/deg when contrast is above 40%. At lower eccentricities, contrast constancy requires less physical contrast and applies over spatial ranges reaching into higher frequencies. These results indicate that all supra-threshold frequencies at any eccentricity are equally perceivable, directly contradicting the threshold-based strategy of treating high frequency and high eccentricity detail as less important.

Sensitivity to supra-threshold stimuli is also increased by summing when those stimuli are complex. Research with complex supra-threshold stimuli is still more sparse [Bonneh and Sagi 1998, 1999], but early results indicate that like sub-threshold summing, supra-threshold summing is linear within a channel. However, while all results indicate that supra-threshold summing across channels is at least sub-linear, some research shows that it is nearly or even supra-linear. Bonneh & Sagi [1998, 1999] report a Minkowski exponent of 3.5 across position channels in highly controlled settings, while Chandler and Hemami [2002] report near-linear summing ($\beta = 1.12$) across orientation channels and supra-linear summing ($\beta = 0.67$) across frequency channels in more natural settings. Regardless of its form, summing occurs supra-threshold, meaning that increasing LOD (adding meaningful high spatial frequencies) can compensate for reduced perceptibility when stimulus contrast is low or stimulus eccentricity is high. Once again this flies in the face of threshold-based strategy, which reduces LOD in the same situations.

## 3 A Primer On Experimental Methods

In this section we provide a very brief introduction to experimental methods and terms. For more in-depth treatments, please see [Elmes et al. 2003; Ferwerda et al. 2002].

Psychologists form *hypotheses* about the human process or task they wish to understand, and then in an experiment manipulate certain *independent variables*, leaving *control variables* unchanged, and observing the *effect* on *dependent variables*. Each independent variable may also be called a *factor*, and each possible value of a factor is a *level*. Two or more experimental factors are usually combined or *crossed*, with each resulting combination called a *treatment*. For example, completely crossing two three-level factors results in nine treatments.

To distinguish between signal and noise in their results, psychologists employ many statistical tools, especially the *analysis of variance* (ANOVA), which estimates the likelihood that the effect of an independent on a dependent variable is only random with the probability $p \, \varepsilon \, [0,1]$, where 1 indicates the effect is certainly random and non-existent. By convention, psychologists call any effect with $p \leq .05$ *significant*, and with $p \leq .1$ *marginally significant*. The related effects of two or more independent variables are called an *interaction*. ANOVAs that are sensitive to such effects of $n$ variables are called *n-way*.

Some of the largest sources of noise in psychological experimentation are the differences between individual experimental participants. When each participant performs the task at all levels of an independent variable, between participant differences are largely sidestepped, the sensitivity or *power* of the analysis can increase, and the variable is said to vary *within subject*. *Repeated measures* ANOVAs exploit within subject variables to increase power by treating the performance of each participant at a variable level as an additional measure of the process being studied, and comparing variability over these measures – over the *random factor* – to variability introduced by the independent variables. To increase sensitivity further, each participant often repeats the experimental task several times in each treatment, and the results are averaged *per participant* (over repetition) before analysis. In such cases, treatment repetition becomes an additional source of noise, and results can instead be averaged *per repetition* (over participants) to compare the effects of independent variables and task repetition.

ANOVAs indicate only whether the overall effect of an independent variable is significant. To determine if the change between two variable levels is significant, psychologists turn to *follow-up analyses*. In this paper we use Bonferroni comparisons.

## 4 Discrimination Experiment

Applied research indicates that peripheral threshold-based LOD control can be quite effective, while basic perceptual research suggests that effectiveness cannot be explained by supra-threshold perceptibility. To investigate this apparent contradiction, we performed two experiments. Our first experiment was designed to study the relationships between supra-threshold *LOD*, *contrast*, *eccentricity*, and efficient location of complex potential targets in the visual periphery during search. In a three alternative forced choice (3AFC) task, participants were asked if a single displayed target was present in the left periphery, present in the right periphery, or simply absent. By introducing uncertainty about target location and requiring identification of that location, we change what might otherwise be a classic test of peripheral detection into a more demanding test of peripheral discrimination. Discrimination of this sort is relied on extensively to guide eye and head motion in more complex visual search tasks.

Threshold-based control makes the following predictions about these relationships. First, *discrimination performance will decline as LOD or contrast drops*, since eliminating perceivable spatial frequencies or reducing perceivable contrast will always reduce target perceptibility. However, *eliminating fine LOD will have less impact than eliminating coarse LOD*. This prediction arises from the CSF, which indicates lower threshold visual sensitivity to high spatial frequencies. (Any reduced sensitivity at the lowest frequencies is generally ignored, since those frequencies are almost always displayed). Next, *performance will decline as eccentricity rises, particularly when contrast is low*. The logic here is that since threshold sensitivity drops at high eccentricity, supra-threshold perceptibility will also drop, especially when contrast is already low. Finally, *LOD will have less impact on performance when eccentricity is high and/or contrast low*, since threshold sensitivity to a given spatial frequency drops as eccentricity increases or contrast drops.

Supra-threshold perceptual research also predicts that *discrimination performance will decline as LOD or contrast drops*, using the same reasoning as threshold-based control. But in addition, it predicts that *eliminating fine LOD will have the same impact as eliminating coarse LOD*, due to contrast constancy. Like threshold-based control, supra-threshold perceptual research anticipates that *performance will decline as eccentricity rises, particularly when contrast is low*. Its reasoning however is that increasing eccentricity reduces the number of channels that respond to the target, especially when contrast is already low. Lastly, supra-threshold research contradicts threshold-based control by predicting that *LOD will have more impact on performance when eccentricity is high and/or contrast low*, because the spatial frequency bandwidth of the target is especially important when the strength of channel response is already weakened by display at high eccentricity or low contrast.

In general, we expected that the predictions of supra-threshold research would be correct. But to account for the success of threshold-based control in applied LOD experimentation, we hypothesized that for tasks such as visual discrimination and search, the importance of a peripheral visual detail is related not to its own perceptibility, but rather to the contribution it makes to overall target perceptibility. As long as target discrimination is not completely reliable, any detail is just as important as another. When collected target detail has made discrimination completely reliable, any additional detail is just as unimportant as any other. We therefore predicted that *all LOD coarser than a certain level of maximum reliability would have equal impact on performance, while all LOD finer than this level would have no impact on performance*. This would explain the effectiveness of the threshold-based approach of eliminating fine LOD first.

## 4.1 Methods

Ten graduate students participated in the experiment. All reported normal vision, none wore eyeglasses.

The experiment used a three factor design, with the independent variables within-subject. These variables were *LOD* (10 x 8, 20 x 15, 30 x 23 and 40 x 30 display pixels), target/background *contrast* (high, low and lowest), and target *eccentricity* (five levels: ±20 degrees from view center, ±30 degrees, and not visible). These variables were completely crossed, giving a total of 4 x 3 x 5 = 60 experimental treatments.

Participants viewed the experimental scene using the Virtual Research Flight Helmet. Though by now a very old device, we chose this head-mounted display (HMD) because it offers a wider field of view (FOV) than almost all more recently produced HMDs. The Flight Helmet weighs 3.7 lbs, and in front of each eye places a color LCD screen with a vertical FOV of 58.4 degrees, and a horizontal FOV of 75.3 degrees. Each screen contains 208 horizontal by 139 vertical pixels, or angular resolution of 1.38 horizontal and 1.19 vertical cycles per degree (c/deg) – more than adequate for our experimental needs. We used the Flight Helmet in a biocular mode by sending the same image to each of the video inputs, and using an optical adjustment provided by Virtual

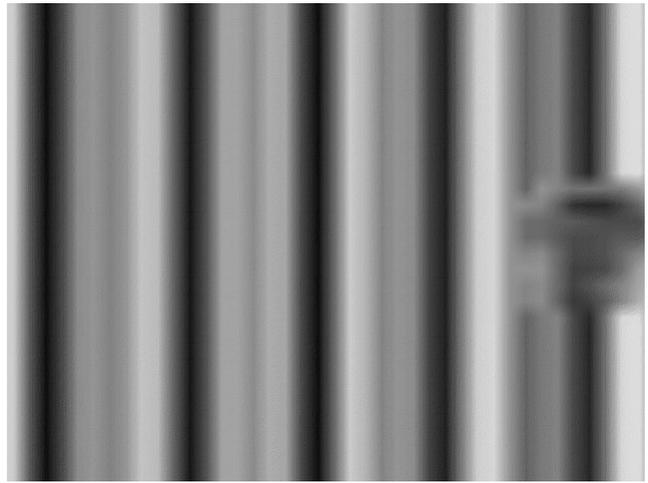

**Figure 1:** A view presented in the first experiment. Here, the target is presented over the lowest *contrast* background at the highest *eccentricity* and the 20 x 15 *LOD*.

Research that enables users to fuse the two resulting images. Head tracking was not used.

The NTSC signal sent to the Flight Helmet was generated by a Silicon Graphics Onyx Reality Engine II using the SVE virtual environments toolkit [Kessler 2000]. *LOD* was controlled in each frame by rendering the current view to a viewport of the appropriate size, and using that viewport as the texture for a 2D polygon in another viewport sized to span NTSC resolution, and sent to the Flight Helmet. Frame rate averaged 29.9 Hz, with a standard deviation of 0.1 Hz.

Participants used a mouse shaped like a pistol grip to respond to the experimental environment. The mouse had two buttons for the thumb mounted on top, and one button for the index finger mounted on the front. The mouse was not tracked. Participants were seated while they performed the experiment.

Since our applied interest was effectiveness of supra-threshold LOD control in complex task related contexts, we used more complex stimuli in our experiments than those typically used in basic psychophysical research.

The displayed environment consisted of the view-spanning background indicated by the current level of target/background *contrast*. When the target was present, it was laid over this background at the peripheral location indicated by *eccentricity*. This entire scene was then displayed at the resolution indicated by *LOD*. The target was a 12 degree square polygon textured with the monochromatic image of a human face. For high *contrast* treatments the background was completely black; the low *contrast* background was a vertical 8-bar gray grating reaching minimum and maximum display luminance in each cycle; lowest *contrast* treatments added a similar 15-bar grating to the 8-bar grating and normalized the result (see Figure 1). If we compare the average luminance of the target ($L_t$) and the three backgrounds ($L_b$) using Michelson contrast ($|L_t-L_b|/(L_t+L_b)$), we obtain 1 for high, 0.055 for low and 0.051 for lowest *contrast*. However, this ignores the many spatial scales of contrast in our complex target and backgrounds. Following Scharff et al. [2000], Chandler and Hemami [2002] suggest relative root mean squared (RMS) contrast for measuring the contrast of a complex target in a complex background. The RMS contrast of an image is $(1/L_{avg})(1/n \, \Sigma_1^n \, (L_i-L_{avg})^2)^{1/2}$, where $L_{avg}$ is the average luminance, $L_i$ the luminance at a pixel, and $n$ the number of pixels in the image. Relative RMS contrast is the ratio of the RMS contrast of the target with its background over the RMS contrast of the target without its background. The relative RMS contrast at the high *contrast* setting is again 1, at low *contrast* it is 0.114, and at lowest it is 0.082. Clearly the primary difference between the low and lowest *contrasts* is spatial content.

Depending on the current *LOD* level, these environments were displayed at the very coarse angular resolutions of .067 x .068, .133 x .128, .199 x .197 and .266 x .257 c/deg. Note that all these levels could resolve at least the lowest frequencies of the target (.042 c/deg) in display, and are perceivable at 30 degrees of eccentricity, where acuity is roughly 5.4 c/deg [Koenderink 1978]. According to Graham [1989], cross channel summing begins when the highest frequency component is three times the frequency of the lowest component. This *LOD* manipulation therefore spans a range of spatial frequencies wide enough to cause summing both within and across frequency channels.

In each trial, the view indicated by the current treatment was presented for 150 ms, inadequate time for any eye motion [van Diepen et al. 1999] (ensuring proper control of target location in the visual field). A blank white screen was then displayed, ensuring that participants could not use a visual afterimage to complete their task. After one second, the white screen was replaced with a blank black screen. Participants responded to the presented view by pressing one of the three joystick buttons to indicate if the target was visible to the left, right, or not at all. The next trial would begin two seconds after the participants responded. If participants responded during the presentation of a view or the blank white screen, the black screen would immediately be presented in preparation for the next trial. Participants could take as long as they wanted to respond.

Participants began the experiment by performing 20 unrecorded practice trials at full HMD resolution. The following trials were organized into 12 randomly ordered blocks, corresponding to the 12 combinations of *LOD* and *contrast*. In each block, participants performed 5 practice trials, followed by 50 recorded trials, corresponding to 10 repetitions at each of the five levels of *eccentricity*. The order of the trials within each block was random. Participants thus performed a total of 680 trials, of which 600 were recorded. Participants were able to take a break between any block. All participants finished the experiment within an hour.

### 4.2 Results

We analyzed the results from the peripheral discrimination task using a three-way (4 *LOD* x 3 *contrast* x 2 *eccentricity*), repeated measures, per-participant ANOVA. The "not visible" level of the *eccentricity* variable was included in the design only for control (ensuring that participants did not simply fixate to one display side) and results from trials using that level were excluded from analysis. The other four levels of *eccentricity* were combined for the purposes of analysis into two levels: ±20 degrees and ±30 degrees. The dependent variable was discrimination accuracy, the percentage of trials in which the target was correctly located.

Table I shows the ANOVA results, which indicate that all possible effects were significant, including all interactions. Increasing *eccentricity* reduced peripheral discrimination accuracy. Follow-up analyses showed that increasing *LOD* continued to improve accuracy significantly until *LOD* moved from its second-highest to its highest level. High target/background *contrast* allowed significantly more accuracy than both the low and lowest *contrast* levels.

Figure 2 graphs the interaction of *contrast* with *eccentricity*, showing means and standard error (as do all graphs). When *contrast* was high, participants almost always found targets, regardless of where they were located. But when *contrast* was lower, participants had more trouble finding targets located at 30 degrees than those at 20 degrees. Figure 3 shows the interaction of *LOD* and *eccentricity*. When targets were located at 20 degrees, accuracy suffered only when *LOD* dropped to its lowest level. On the other hand, when targets were located at 30 degrees, accuracy suffered also when *LOD* dropped to the second lowest level. Figure 4 illustrates the interaction of *LOD* and *contrast*. When *contrast* was high, *LOD* had little effect on accuracy; but when *contrast* was lower, decreasing *LOD* caused steep drops in accuracy, ultimately reaching chance levels. Finally, Figure 5 shows the interaction of *LOD*, *contrast* and *eccentricity*. *LOD* and *eccentricity* only interacted when participants searched with lower *contrast*. When *contrast* was high, participants were always able to find the target reliably.

Table I: ANOVA Results For First Experiment

| Independent Variable | ANOVA |
|---|---|
| LOD | $F(3,27) = 101.23$, $p < .00005$ |
| Contrast | $F(2,18) = 86.94$, $p < .00005$ |
| Eccentricity | $F(1,9) = 101.90$, $p < .00005$ |
| LOD * Contrast | $F(6,54) = 47.57$, $p < .00005$ |
| LOD * Eccentricity | $F(3,27) = 52.47$, $p < .00005$ |
| Contrast * Eccentricity | $F(2,18) = 24.06$, $p < .00005$ |
| LOD * Eccentricity * Contrast | $F(6,54) = 8.76$, $p < .00005$ |

### 4.3 Discussion

We begin our discussion of these results with some cautionary notes, move on to the results themselves, and conclude with a review of some of the experiment's immediate implications. We will discuss wider implications at the end of this paper.

Because peripheral visual sensitivity is quite poor, we used very coarse LOD manipulations and a relatively large target. Experimentation in this spatial range also ensured the capture of any supra-threshold LOD effects. Even this range was enough to capture the full range of performance, from chance to nearly perfect accuracy. Should users be seeking smaller targets, we would expect that improved LOD would be required to maintain discrimination accuracy – once more however, these expectations should be confirmed directly.

Our applied interests led us to focus on the effects of relative contrast rather than absolute contrast. Future efforts might ensure that contrast levels are spaced more evenly and finely, with luminance measured more precisely using light meters. In addition, the backgrounds we used to control contrast were spatially very simple, containing just zero, one or two horizontal spatial frequencies. These are certainly not typical of applied settings, and this work might be strengthened with a reexamination using more complex backgrounds.

We approximated applied LOD management using a render-to-texture technique that provided good control of spatial frequency, but eliminated many of the high frequencies at polygon edges that would be present if dynamic LOD management were used. Most of these frequencies are an unintended byproduct of simplification, however when they are (often purposefully) aligned with large model contours and silhouettes, so they may indeed be beneficial. Further work might examine this very practical question directly.

Reassuringly, results indicated that most of the experimental *LOD* levels must be supra-threshold. As Figure 5 makes clear, only three treatments using the 10 x 8 *LOD* and lower *contrast* approached chance levels of accuracy. (Since the 10 x 8 *LOD* was well above acuity thresholds, this likely indicates an increase in threshold for our task due to reduced *contrast* and the need to discriminate target direction). Any increase in *LOD* brought an improvement in accuracy if it was not already near maximum, indicating clearly that the additional detail was perceivable.

Results were quite consistent with our hypotheses. The identical predictions of threshold-based control and supra-threshold perceptual research were correct. Reduced *LOD* or *contrast* caused significant harm to peripheral discrimination accuracy, as did increased target *eccentricity* (when *contrast* was lower). Threshold-based and supra-threshold predictions about the relationships of *LOD* to *contrast* and *eccentricity* differed, and as we expected results followed the supra-threshold predictions: *LOD* became more important as *contrast* declined and as *eccentricity* rose. Finally, according to supra-threshold research all perceivable *LODs* are equally important, while threshold-based control assumes that fine *LOD* is always less important than coarse *LOD*. We speculated that below a certain *LOD* level achieving reliable target discrimination, all *LODs* would have equal impact on accuracy, while above this level, all *LODs* would have no impact. Our results fit this pattern almost perfectly.

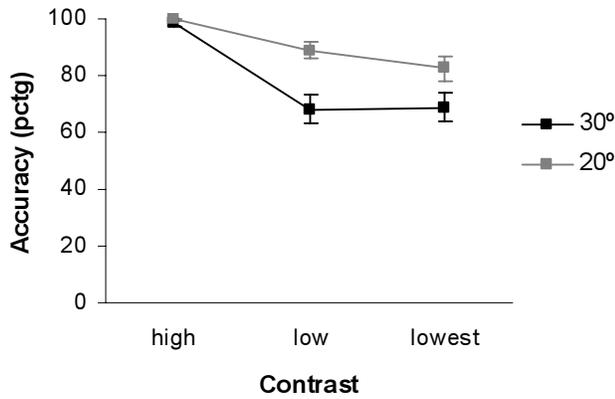

Figure 2: The effects of *eccentricity* and *contrast* on discrimination accuracy. E*ccentricity* is shown by gray level, *contrast* by the abscissa, and percent correctly located targets averaged over *LOD* by the ordinate.

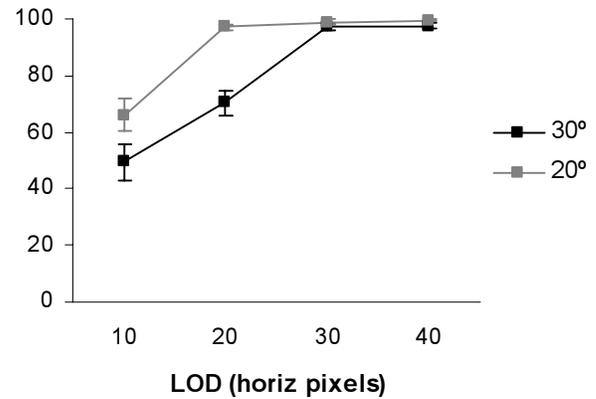

Figure 3: The effects of *LOD* and *eccentricity* on discrimination accuracy. *Eccentricity* is shown by gray level, *LOD* in horizontal pixels by the abscissa, percent correctly located targets averaged over *contrast* by the ordinate.

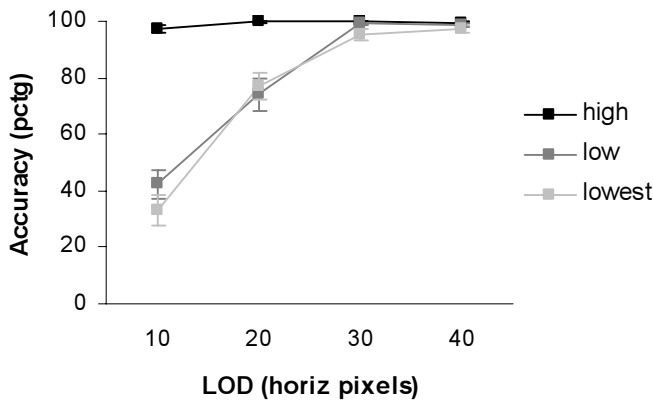

Figure 4: The effects of *LOD* and *contrast* on discrimination accuracy. *Contrast* is shown by gray level, *LOD* in horizontal pixels by the abscissa, percent correctly located targets averaged over *eccentricity* by the ordinate.

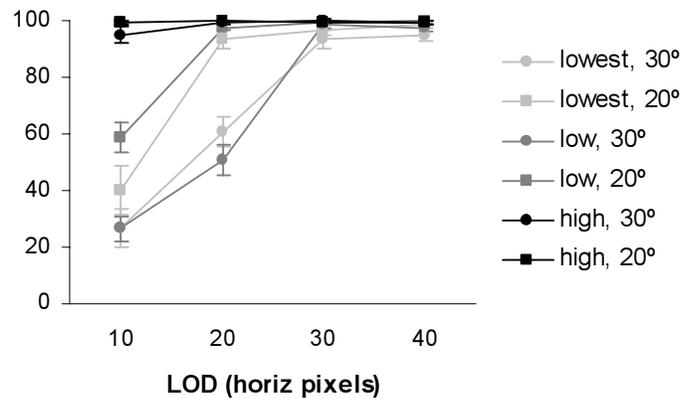

Figure 5: The related effects of *LOD*, *eccentricity* and *contrast* on discrimination accuracy. *Contrast* is shown by gray level, *eccentricity* by point shape. *LOD* in horizontal pixels is on the abscissa, percent correctly located targets is on the ordinate.

The assumption that threshold-based methods can be used for supra-threshold LOD control is dangerous. In particular, threshold-based control reduces LOD more aggressively when eccentricity is high or contrast low. Our results indicate that LOD should sometimes be *increased* in these difficult settings, compensating for the challenging environment and preserving a reliable level of perceptibility. Threshold-based control also eliminates fine LOD before coarse LOD, since threshold sensitivity to fine LOD is lower than threshold sensitivity to coarse LOD. Our results indicate that this control strategy can be effective, but support an alternative explanation for its effectiveness. In particular, LOD control in the periphery need only support a fully reliable level of *object* perceptibility. Below that level, all scales of detail are equally important. Above that level, all scales of detail are equally unimportant.

## 5 Search Experiment

Our first experiment provided surprising evidence that called threshold-based supra-threshold LOD control into question, and provided a potential explanation for its effectiveness in visual search tasks. Yet the experiment examined LOD control using discrimination and not search, measured performance using only accuracy and not time, and did not actually implement variable LOD control. In our second experiment, we addressed these concerns and examined the first experiment's conclusions further using methods much like those we used to demonstrate threshold-based LOD's effectiveness in [1997a, 1997b]. Participants sought one target among four distractors in a random location, using a head-tracked display as peripheral *LOD* and *contrast* were varied. *LOD* included the same four levels used in the first experiment, as well as two new control levels included for comparison: a blank level in which nothing was displayed in the periphery, and a full level in which the central and peripheral display regions used the same full resolution *LOD*. *Contrast* was varied by using only the high and lowest levels from the first experiment, since results of that experiment showed that the low and lowest levels had identical effects. Eccentricity was not a variable in this experiment, since *LOD* was only varied at high eccentricities.

## 5.1 Motion, Search and Implications

There are several important differences between this experiment and the first. The head-tracked experimental setting introduces motion, a very complex phenomenon surveyed in [Smith and Snowden 1994] and [Sekuler, Watamanuik and Blake 2002]. Research on contrast sensitivity thresholds during motion [Kelly 1979; Koenderink et al. 1978; Watson and Ahumada 1985; Watson and Turano 1995] consistently shows that sensitivity is reduced by motion, with the CSF being translated toward lower spatial frequencies and acuity dropping fairly rapidly as velocity increases. When moving stimuli are in the visual periphery, sensitivity is already lowered by eccentricity, and velocities must be higher before they reduce sensitivity [Graham 1989; Kelly 1984; Koenderink et al. 1978]. Research on supra-threshold contrast sensitivity during motion is just beginning, with early results indicating that sensitivity to rapid change improves as supra-threshold contrast increases [Fredericksen and Hess 1997], and that this effect depends on spatial frequency [Fredericksen and Hess 1999].

Contrast sensitivity is just one measure of motion perception, two of many others are speed discrimination and minimum motion thresholds. Speed discrimination of quickly moving targets (30 deg/sec and above) is constant throughout the visual field, but ability to discriminate between slower speeds drops quickly outside the fovea [van de Grind et al. 1986, 2000; McKee and Nakayama 1984]. Speed discrimination is unaffected by variation in contrast above 5% [McKee et al. 1986; Turano and Pantle 1989]. Minimum motion thresholds measure the smallest detectable distance traversed. As motion moves into the periphery, these thresholds rise [Tynan and Sekuler 1982, McKee and Nakayama 1984]. Contrast has little effect on motion thresholds of single-frequency targets [Nakayama and Silverman 1985], but increases in contrast up to roughly 50% improve the motion thresholds for complex targets.

This experiment also uses a visual search task, which has a long history of study surveyed in [Rayner 1998] and [Liversedge and Findlay 2000]. Early study of search focused on the role of attention [Treisman and Gelade 1980], proposing that in an initial pre-attentive stage basic target features are registered in parallel (and thus "pop out"), and in a second stage attention is required for further serial processing of targets. More recent research has revealed that basic perceptual factors play a larger role than previously thought. In particular, moving targets into the visual periphery, eliminating spatial frequencies and reducing target contrast consistently harms search performance [Carrasco and Frieder 1997; Geisler and Chou 95; Notdurft 2002]. Targets moving in fields of relatively slow or static distractors are quite easy to find [Royden et al. 2001], while performing any search in a dynamic environment is more difficult than in a static setting [van Loon et al. 2003; Niemann and Hoffman 1997; Royden, et al. 2001]. The visual periphery plays an important role in guiding the next saccade, but can be degraded far past eccentric acuity levels without harming performance [van Diepen and Wampers 1998; Parkhurst and Niebur 2002; Saida and Ikeda 1979; Shiori and Ikeda 1989].

In balance, this research suggests that reducing *LOD* in the visual periphery will prove effective for our search task. Performance should improve as *LOD* and *contrast* rise, though due to the effects of global motion it may take longer to reach its peak. Some readers may be surprised to learn that the human visual system is not *more* sensitive to motion in the periphery – as we have just pointed out, isolated moving objects are quite salient in search. Graham [1989] and Anstis [1986] suggest that this common misunderstanding may be caused by the fact that in relative terms, eccentric motion sensitivity is often superior to eccentric static sensitivity: for example temporal acuity and high-velocity speed discrimination are *unchanged* (not improved) in the visual periphery, while spatial acuity drops steeply.

Therefore despite the new elements of motion and search in this experiment, our hypotheses for this experiment echoed those of the first. We again expected that *performance would be harmed when contrast or LOD was reduced*, as predicted by both threshold-based control and supra-threshold perceptual research. We also anticipated that *the amount of LOD required to maintain performance would be increased when contrast was reduced*, contradicting threshold-based control. While the first experiment indicated that the level of maximum reliability was reached at *LOD* 30 x 23, based on the above review of research we hypothesized cautiously that *all LOD coarser than 40 x 30 would have equal impact on performance, while all LOD above that level of maximum reliability would have no impact on performance*. An *LOD* of 40 x 23 would therefore be equivalent to full *LOD*, and would allow participants to guide head motion reliably so that candidate targets could efficiently be brought into the full *LOD* central display region for further examination.

## 5.2 Methods

Ten graduate students participated in the experiment. Of these, seven had also participated in the first experiment. All reported normal vision, none wore eyeglasses.

The experiment used a two-factor design, with the independent variables within-subject. These variables were peripheral *LOD* (blank, 10 x 8, 20 x 15, 30 x 23, 40 x 30 and full) and *contrast* (high and lowest). These variables were completely crossed, giving a total of 6 x 2 = 12 experimental treatments.

Experimental equipment and setup was largely identical to the first experiment. We note only the differences from that experiment here. The motion of a participant's head in this experiment was tracked with the Polhemus Isotrak II 3D tracking hardware. Participants were free to turn and move their heads; all of this motion was represented in the displayed environment. Display always contained a 30 x 30 degree full *LOD* central region, surrounded by a peripheral region with the resolution indicated by the current *LOD* level. The display was generated in each frame by rendering the current view to two undisplayed viewports, and combining those viewport images dynamically into a single displayed view. One viewport corresponded to the central display region and was always rendered at the maximum HMD resolution. The other peripheral viewport was rendered at a resolution corresponding to the current level of *LOD*. Each of these views was dynamically textured onto a 2D image-space polygon in the displayed view, with the peripheral polygon spanning the entire HMD view, and the full *LOD* polygon spanning the central 30 x 30 degrees of the HMD, occluding the other and being alpha blended with the other at its boundaries (softening the central/peripheral transition, see figures 6 and 7). Earlier we showed that head- (not eye-) tracked 30 x 30 degree full central views support head-tracked search just as well as full detail display in a very similar experiment [1997a].

Frame rate in this experiment was limited to a maximum of 12 Hz, the highest of the minimum frame rates encountered using this unusual rendering technique across all levels of *LOD*. Frame rate averaged 11.96 Hz, standard deviation was 0.12 Hz. The corresponding mean frame time was 83.33 ms. We enforced this limit by introducing a delay into the main display loop whenever a particular frame was too fast. System responsiveness (lag between physical motion and corresponding display motion) was 260 ms.

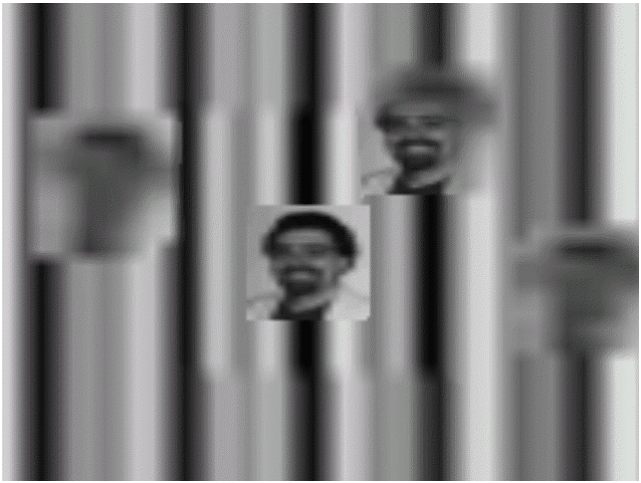 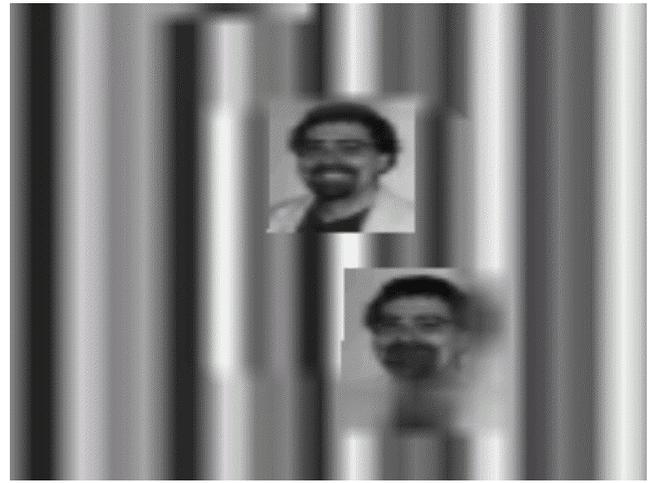

**Figure 6:** A view presented in the second experiment. Here the periphery uses the 20 x 15 *LOD*, while the lowest *contrast* background is used. The central area is (always) displayed at the highest HMD resolution. Four distractors are shown.

**Figure 7:** Another view presented in the second experiment. Again the periphery uses the 20 x 15 *LOD* and the lowest *contrast* background is used. Two distractors (mouth open) and the target (mouth closed) are shown.

Participants were standing inside a 4x4 foot platform while they performed the experiment. The platform was raised six inches off of the floor and surrounded by a 3 foot railing, which kept participants within 4 feet of the Isotrak transmitter.

*Stimuli.* In this experiment the target object was located among four distractors. Both target and distractors were rectangular objects of the same size as the target in the first experiment. The target was textured with the face from the first experiment, but the mouth was closed. All distractors were textured with the same face, with the mouth open. The difference between the faces was thus a feature spanning roughly 2-3 degrees in the horizontal, 1-2 in the vertical – easily resolved by HMD resolution at the display center in both dimensions (compare figures 6 and 7). The target and distractors were clustered and located randomly in a search space 150 horizontal by 118 vertical degrees in size (twice the HMD's horizontal and vertical FOV). Each cluster fit completely into a single HMD view, with target and distractors located randomly in the view.

The displayed environment consisted of a home object (a pedestal textured with a bull's-eye pattern) onto which participants had to focus before each experimental trial. No target or distractor was ever visible when the home object was in the center of the participant's view. The search space was always located to the right of this home position. Behind the home object and the target and distractors was the view-spanning background indicated by the current level of target/background *contrast*, wrapped completely around the participant, repeated and located to ensure that it contained the same spatial frequencies as in the first experiment. This entire scene was then displayed at the peripheral resolution indicated by *LOD*.

During each trial, participants searched for the target using head motion, without ever changing their location in the experimental environment. Participants began each trial by focusing on the home object, and pressing a button to begin. After a random (between .1 and .8 seconds) delay, the home object disappeared, and the target and distractor objects appeared outside the participant's initial view. Participants looked for the target object and pressed one of two buttons to indicate if the object was present or not present. The target and distractors then disappeared, and the home object reappeared. At the same time, onscreen feedback appeared, indicating if the correct button had been pressed, and the number of seconds required for the search. When participants had again focused on the home object and pressed the appropriate button, a new search trial began.

Participants pressed the left thumb button to indicate a target was present, the index button if it was absent. In trials with a target, participants were not credited with a correct trial unless they had brought the target into their view. This eliminated simple guessing as a search strategy. In target absent trials, participants were not credited with a correct trial unless they had brought every object into their view. This forced an exhaustive search.

Before beginning the experiment, participants read a two-page introduction to the purpose of the experiment and its procedure. This explained, among other things, that participants were permitted to pause between any two trials if they required a rest. It also made participants aware that they would be ranked by search time and accuracy, with the participant with the best cumulative ranking receiving $50 after the completion of the experiment.

At the beginning of the experiment, participants performed 20 unrecorded search trials as practice, using the lowest *contrast* and the best *LOD*. Remaining trials were organized into 12 randomly ordered blocks, corresponding to the 12 treatments. Before beginning recorded searches with a new block, participants performed five unrecorded practice search trials, so that they might familiarize themselves with the new display configuration. Within each block, 10 trials with target present and 10 with target absent were ordered randomly. Participants were required to perform all these 20 trials correctly. Any trial incorrectly performed reduced accuracy, and was randomly reinserted into the present/absent ordering of remaining trails and performed again. Participants thus performed a total of 240 correct trials. Including unrecorded practice trials, participants performed at least 320 trials. Most participants performed two experimental sessions, each 45 minutes long. Participants were required to complete all trials for a given block before ending a session.

### 5.3 Results

Because the effects in this experiment were less consistent than in the first, we used both per-participant and per-repetition analyses. We used eight two-way repeated measures ANOVAs (target present or absent, measured with time or accuracy, per-participant or per-repetition, as *LOD* and *contrast* were varied). The dependent variables were search accuracy, the percentage of recorded trials in which the target was correctly identified as present or absent; and search time, the mean time in correctly performed trials from the disappearance of the home object until the participant's trial-ending button press. Table II shows all ANOVA results, while Figures 8 and 9 graph mean and standard error of search accuracy and time for each treatment.

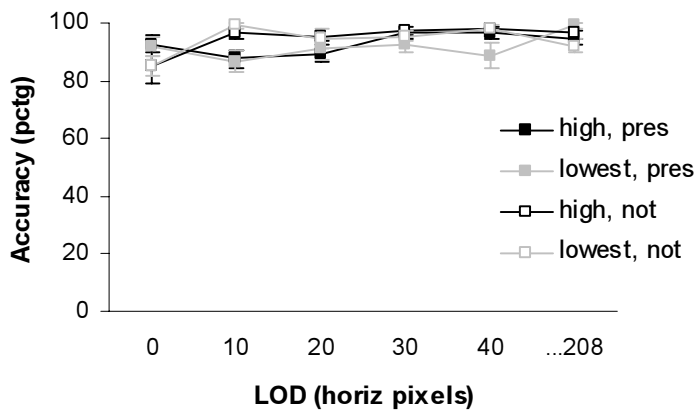
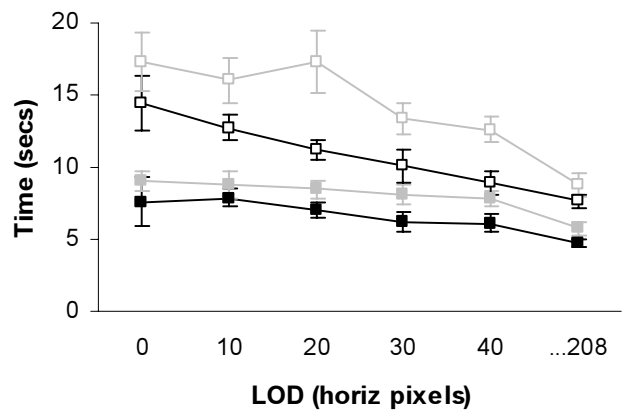

**Figure 8:** Search accuracy as affected by *LOD* and *contrast* in the second experiment. *Contrast* is indicated by curve gray level, target present or absent by point solidity. *LOD* in horizontal pixels is on the abscissa, percentage correctly located targets is on the ordinate.

**Figure 9:** Search time as affected by *LOD* and *contrast* in the second experiment. *Contrast* is indicated by curve gray level, target present or absent by point solidity. *LOD* in horizontal pixels is on the abscissa, search time is on the ordinate.

Table II. ANOVA Results For Second Experiment

| Indep Var | Tgt Pres | Dep Meas | Per | ANOVA |
|---|---|---|---|---|
| *LOD* | yes | accuracy | participant | $F(5,45) = 3.11, p = .017$ |
| *LOD* | " | " | repetition | $F(5,45) = 3.77, p = .006$ |
| *LOD* | " | time | participant | $F(5,45) = 6.72, p < .0005$ |
| *LOD* | " | " | repetition | $F(5,45) = 19.08, p < .0005$ |
| *LOD* | no | accuracy | participant | $F(5,45) = 6.12, p < .0005$ |
| *LOD* | " | " | repetition | $F(5,45) = 8.16, p < .0005$ |
| *LOD* | " | time | participant | $F(5,45) = 18.02, p < .0005$ |
| *LOD* | " | " | repetition | $F(5,45) = 70.11, p < .0005$ |
| *Contrast* | yes | time | participant | $F(1,9) = 32.10, p < .0005$ |
| *Contrast* | " | " | repetition | $F(1,9) = 58.91, p < .0005$ |
| *Contrast* | no | time | participant | $F(1,9) = 35.95, p < .0005$ |
| *Contrast* | " | " | repetition | $F(1,9) = 88.09, p < .0005$ |
| *LOD * Contrast* | yes | accuracy | participant | *$F(5,45) = 2.27, p = .064$* |
| *LOD * Contrast* | no | time | participant | *$F(5,45) = 2.31, p = .059$* |
| *LOD * Contrast* | " | " | repetition | $F(5,45) = 3.73, p = .007$ |

The effects of *LOD* and *contrast* on search accuracy were limited, with participants able to maintain high levels of accuracy in search across all conditions. *Contrast* had no effect on search accuracy at all. *LOD* affected search accuracy in both the per-repetition and the per-participant analyses, but follow-up analyses revealed that these effects were largely restricted to control levels included for comparison. When the target was absent, accuracy with the blank *LOD* was significantly worse than accuracy with all other *LOD*s. When the target was present, accuracy with the full *LOD* was significantly better than accuracy with the 10 x 8 *LOD*. The interaction of *LOD* and *contrast* when the target was present was marginally significant in the per-participant analysis, with high *LOD*s bringing a minor improvement in accuracy when *contrast* was high, but not when *contrast* was low.

Effects on search time were more striking and meaningful, with improved *contrast* or *LOD* significantly reducing search time whether the target was present or absent, according to both analyses. Search times using *LOD* levels with reduced peripheral detail were often better than those with the completely blank *LOD*, but were always worse than those with full *LOD*. The interaction of *LOD* and *contrast* when the target was absent was significant in the per-repetition analysis and marginally significant in the per-participant analysis, with finer *LOD*s bringing steady improvement in search times when *contrast* was high, but only high *LOD*s bringing improvement when *contrast* was low.

### 5.4 Discussion

We confine our discussion largely to the observed effects on search time, since the full detail central display region allowed high accuracy in all conditions. Once more we begin by listing reasons for caution in interpreting the results.

Since this experiment followed up the first and used many of the same procedures and stimuli, most of the limitations we listed for the first are relevant for the second. Here we focus on the differences between this experiment and the first.

Frame times were fairly low in this experiment (12 Hz, 83.33 ms/frame), raising the issue of relevance in systems demanding better interactive response. Certainly these results would be strengthened if a similar study could be performed at a higher mean frame rate (e.g. 30 Hz, 33.33 ms/frame). However, what research exists suggests that results in such a study would not be meaningfully different from ours. We [1998] investigated the effects of both frame rate and resulting delay and found little improvement in imprecise, ballistic task performance (such as the head motion in this experiment) as frame rate rose above 12 Hz.

Different target-distractor configuration can affect search performance in a consistent manner, and might conceivably interact with *LOD* and *contrast*. In [1997b], we found that fewer distractors, reduced mean target distance from the starting search position, and tighter distractor clustering all reduce search times. However, they did not interact with supra-threshold *LOD*. This would not lead us to expect a relationship between *LOD* or *contrast* and target-distractor configuration in this experiment, but a confirming study would certainly be helpful.

As in the first experiment, increasing *LOD* or *contrast* improved performance. *LOD* became more important to participants as *contrast* dropped, contradicting again the predictions of threshold-based control. In particular, when the target was absent and *contrast* high, every *LOD* level improved performance. But when target was absent and *contrast* at its lowest, performance did not begin to improve until *LOD* passed the 20 x 15 level. This level of minimum reliability is the level at which *LOD* began to improve detection of potential targets beyond the blank *LOD* level.

Generally, equal improvements in *LOD* brought equal improvements in performance, in agreement with the predictions of supra-threshold perceptual research. However, an unexpected level of minimum reliability was found, and since the 40 x 30 level was never as reliable as full *LOD,* the level of maximum reliability was never reached.

This experiment confirms the danger of basing supra-threshold LOD control on perception at threshold. In comparison to the first, this experiment's task was more dynamic and higher level, and its display more like those used in the field. Despite these differences, threshold-based predictions were once more contradicted. Rather than needing less detail when contrast was low, participants required more. Rather than finding coarse detail more useful than fine detail, participants found them equally useful. Finally, while in the first experiment *LOD* reached a level of maximum reliability and threshold-based control was coincidentally effective, in this experiment maximum reliability was never reached (except in control).

These results also illustrate the limitations of basing display in applied settings on knowledge of performance in low-level perceptual tasks. We based many of this experiment's hypotheses on such knowledge, and overall they proved to be fairly accurate. But despite performing a low-level discrimination experiment with stimuli very similar to those in this search experiment, we were unable to predict the level of maximum reliability, nor did we anticipate a level of minimum reliability. Consider for example the lowest *contrast* target absent and target present conditions. When the target was absent a level of minimum reliability was found, while when the target was present no such level was found. Since the only displayed difference between these two conditions was a very small feature (the mouth), higher-level cognitive factors such as decision criteria likely account for most of the large differences in the utility of improved *LOD*.

## 6 Applied Implications

These experiments support several immediate recommendations for those managing supra-threshold peripheral LOD in the field. Naturally, applied contexts will differ from our experimental setting, and our guidelines should be viewed with the same well-informed and critical eye application designers use to evaluate any potential element of their unique systems.

*Threshold-based LOD control can still be used when supra-threshold contrast is low.* The transition to constancy does not occur until contrast is at least 10%, with greater contrast required at higher eccentricities. Until this transition occurs, perceptibility drops as eccentricity increases, and its dependence on contrast and spatial frequency is fairly accurately described by the CSF. Thus fine detail should be eliminated before coarse, low contrast detail before high, and eccentric detail before foveal.

*But LOD control at higher contrasts should be based on contrast constant perceptibility.* After the transition to constancy occurs, perceptibility continues to decline as eccentricity increases, but spatial frequency is a less effective predictor of perceptibility. Thus low contrast detail should be eliminated before high, and eccentric detail before foveal.

*When LOD control begins to affect task performance, detail should be preserved where sensitivity is lowest.* When LOD control affects task performance, it is because the user lacks required information. Information will be sparsest where sensitivity and thus display reliability is lowest. Thus LOD control strategy should be reversed: detail should be *added* to low contrast regions before high, and to eccentric regions before foveal. This may imply removing and redistributing detail from high contrast or less eccentric display regions. Fortunately, this complicating reversal of control strategy may be avoidable: the amount of detail required to achieve reliability is surprisingly small and far above threshold. Our experimental tasks made heavy use of the periphery, yet the discrimination task required only 0.2 c/deg to maximize performance, while trends indicated 1 c/deg would be reliable for the search task.

## 7 Research Implications

These experiments also generate important questions:

*How much supra-threshold peripheral LOD can be eliminated before task performance is harmed?* The answer will depend on both the low and high-level elements of the user's task. For example in search, the spatial frequency content of the target and background play a large role. If instead the user is only trying to become familiar with an environment, attention will be much less directed, and LOD control may have to be more conservative. Many perceptual scientists have turned to signal detection theory in forming models of supra-threshold detection; it may also prove useful in conservatively estimating reliable supra-threshold LOD.

*What are the implications of this research for supra-threshold LOD control in the fovea?* This research casts great doubt upon threshold-based solutions for supra-threshold LOD control in the fovea. Our results certainly contradicted threshold-based LOD control strategy in the visual periphery, and perceptual research leads us to expect that these trends will only strengthen in the fovea. In fact, most research on supra-threshold perceptual sensitivity probed foveal perception, where contrast constancy is not limited by any decline in acuity due to eccentricity.

*How should the low-level demands of perceptibility be balanced with the high-level demands of task?* By definition, all supra-threshold detail is perceivable, and therefore it is information available to the user. In many cases, information importance and information perceptibility will not correlate. For example when compressing images, Chandler and Hemami [2003] found it useful to supplement contrast constancy with a model of higher level scale-space effects. Similarly, model simplification works despite eliminating supra-threshold detail in fine to coarse order, likely also due to high-level effects. Yet interestingly, Sorkine et al. [2003] have found success with lossy geometry compression that introduces error in coarse rather than fine detail.

## 8 Acknowledgements

Thanks to Alinda Friedman for her statistical knowledge, and to the former members of Georgia Tech's VE group for logistical help. Polymath Jack Tumblin provided many useful leads. This research was supported by NSF grant IIS-0093172.